\documentclass[twocolumn,amsmath,amssymb,nofootinbib,preprintnumbers,showpacs]{revtex4}

\usepackage[pdftex]{graphicx}
\usepackage{epstopdf}

\usepackage{amsmath}
\usepackage{amssymb}
\usepackage{subfigure}
\usepackage{hyperref}
\usepackage{url}
\usepackage{xcolor}
\usepackage{color}
\definecolor{amaranth}{rgb}{0.9, 0.17, 0.31}
\definecolor{purple(munsell)}{rgb}{0.62, 0.0, 0.77}
\definecolor{americanrose}{rgb}{1.0, 0.01, 0.24}
\definecolor{palatinateblue}{rgb}{0.15, 0.23, 0.89}
\definecolor{royalblue(web)}{rgb}{0.25, 0.41, 0.88}
\definecolor{hanpurple}{rgb}{0.32, 0.09, 0.98}
\definecolor{beaublue}{rgb}{0.74, 0.83, 0.9}
\definecolor{carminered}{rgb}{1.0, 0.0, 0.22}
\definecolor{brightpink}{rgb}{1.0, 0.0, 0.5}
\hypersetup{ linktoc=all,
    colorlinks, linkcolor={palatinateblue},
    citecolor={brightpink}, urlcolor={palatinateblue}
}

\def\sideremark#1{\ifvmode\leavevmode\fi\vadjust{\vbox to0pt{\vss
 \hbox to 0pt{\hskip\hsize\hskip1em
 \vbox{\hsize2cm\tiny\raggedright\pretolerance10000
 \noindent #1\hfill}\hss}\vbox to8pt{\vfil}\vss}}}%
\begin{document}
\preprint{NORDITA-2015-130, Alberta Thy 17-15, YITP-15-98}
   \title{Naked Black Hole Firewalls}
   \author{Pisin Chen$^{1,2}$\footnote{pisinchen@phys.ntu.edu.tw},
   Yen Chin Ong$^{3}$\footnote{yenchin.ong@nordita.org},
   Don N. Page$^{4}$\footnote{profdonpage@gmail.com},
   Misao Sasaki$^{5}$\footnote{misao@yukawa.kyoto-u.ac.jp} and
   Dong-han Yeom$^{1}$\footnote{innocent.yeom@gmail.com}
   }

   \affiliation{%
   ~\\
$^{1}$Leung Center for Cosmology and Particle Astrophysics, National Taiwan University, Taipei 10617, Taiwan\\
$^{2}$Kavli Institute for Particle Astrophysics and Cosmology, SLAC National Accelerator Laboratory, Stanford University, CA 94305, U.S.A.\\
$^{3}$Nordita, KTH Royal Institute of Technology \& Stockholm University, Roslagstullsbacken 23, SE-106 91 Stockholm, Sweden\\
$^{4}$Theoretical Physics Institute, University of Alberta, Edmonton, Alberta T6G 2E1, Canada\\
$^{5}$Yukawa Institute for Theoretical Physics, Kyoto University, Kyoto 606-8502, Japan\\
}%


\begin{abstract}
In the firewall proposal, it is assumed that the firewall lies near the event horizon and should not be observable except by infalling observers, who are presumably terminated at the firewall.  However, if the firewall is located near where the horizon would have been, based on the spacetime evolution up to that time, later quantum fluctuations of the Hawking emission rate can cause the ``teleological'' event horizon to have migrated to the inside of the firewall location, rendering the firewall naked.  In principle, the firewall can be arbitrarily far outside the horizon.  This casts doubt about the notion that firewalls are the ``most conservative'' solution to the information loss paradox.   
\end{abstract}

\date{March 8, 2016}

\maketitle


The black hole information loss paradox \cite{hawking1976} is still unresolved almost 40 years after the issue was raised by Hawking. The debate was further heated by the firewall proposal raised by Almheiri, Marolf, Polchinski and Sully (hereinafter, AMPS) \cite{amps}. See also AMPSS (short for AMPS, together with Stanford) for further arguments and clarifications \cite{apologia}. Essentially, AMPS pointed out that local quantum field theory, unitarity, and no-drama (the assumption that infalling observers should not experience anything unusual at the event horizon if the black hole is sufficiently large) cannot all be consistent with each other. Implicitly, it is also assumed that the Bekenstein-Hawking entropy corresponds to the statistical entropy of the black hole, which not everyone agrees --- see \cite{coy, Giddings1} for recent reviews. Furthermore, it is assumed that there exists an observer who could collect all the Hawking radiation so as to attempt to violate the no-cloning theorem of quantum information by eventually falling into the black hole. AMPS argued that the ``most conservative'' resolution to this inherent inconsistency between the various assumptions (hereinafter, the AMPS paradox) is to give up no-drama. Instead, an infalling observer would be terminated once he or she hits the so-called firewall. This seems rather surprising because the curvature is negligibly small at the event horizon of a sufficiently large black hole,  and thus one would expect nothing special but low energy physics.  

Essentially, the argument for a firewall is the following. Assuming unitarity, the information contained inside a black hole should eventually be recovered from the Hawking radiation. The information content is presumably encoded in the highly entangled Hawking radiation, and it is usually argued that the information should start to ``leak out'' after the black hole has lost approximately half of its initial Bekenstein-Hawking entropy, at the Page time \cite{page1, page1b, page2}. A black hole that has passed its Page time is said to be ``old''; otherwise the black hole is considered ``young.''
In other words, the late time radiation purifies the earlier radiation (which was emitted before the Page time and is --- to a very good approximation --- thermal). Thus, as the AMPS argument goes, the late time radiation is maximally entangled with the earlier radiation; and by the monogamy of quantum entanglement, the late time radiation cannot be maximally entangled with the interior of the black hole. This means that the field configuration across the event horizon is generically not continuous, which leads to a divergent local energy density. More explicitly, we recall that the quantum field Hamiltonian contains terms like $(\partial_r \varphi)^2$. The derivative is divergent at some $r=R$ if the field configuration is not continuous across $R$. This is the firewall. (See also \cite{sam}, and p.\ 26 of \cite{Giddings2}.) 

Usually it is thought that firewalls lie on the black hole event horizons. Of course in quantum mechanics there are no sharp boundaries, and the positions of event horizons should be uncertain, up to perhaps fluctuations of the order of the Planck length. That is to say, firewalls are presumably like stretched horizons \cite{STU}, with the crucial difference that anything that hits a firewall gets incinerated instead of just passing right through, unharmed \cite{apologia}. It is also possible that firewalls lie slightly \emph{inside} the event horizons. In that case a firewall would fall toward the (assumed spacelike) singularity (or whatever replaces the singularity in the quantum theory of gravity) faster than the black hole could shrink in size. However, supposedly a new firewall will be dynamically ``replenished'' on each fast-scrambling time scale\footnote{We shall restrict our attention to the asymptotically flat 4-dimensional Schwarzschild black hole. The fast-scrambling time is of the order $M \log M$ \cite{HP}, c.f. the information retention time, which is of the order $M^3$.} \cite{apologia}. By the nature of the event horizon, if the firewall lies either inside or exactly on the horizon, then it is completely invisible to the observers outside. For firewalls that are not too far outside the event horizons, it is still doubtful that they are perceptible to far-away observers, since it would seem that such firewalls are well hidden inside the Planckian region of the \emph{local} thermal atmosphere\footnote{The Hawking temperature is a quantity measured at infinity, but the local temperature near the horizon is enormously blueshifted to a trans-Planckian temperature, following the Tolman law. See, e.g., \cite{wald}.}.
 
Here we make the assumption that a firewall, if it exists, has a location determined by the past history of the Hawking evaporating black hole spacetime and is near where the event horizon would be if the evaporation rate were smooth, without quantum fluctuations.  (If the firewall were far inside the event horizon, it would not resolve the paradox that it is proposed to resolve.)  Then we show that quantum fluctuations of the evaporation rate in the future can migrate the event horizon to the inside of the firewall location, rendering it naked.

For simplicity, we shall approximate the metric near the horizon of an evaporating black hole by the Vaidya metric with a negative energy influx:
\begin{eqnarray}
ds^{2} = - \left( 1 - \frac{2M(v)}{r} \right) dv^{2} + 2 dv dr + r^{2} d\Omega^{2}.
\label{Vaidya}
\end{eqnarray}
Here $M(v)$ is the mass of the black hole, which is decreasing as a function of the advanced time $v$.  For a smooth evaporation rate of a spherical black hole emitting mainly photons and gravitons, we shall take (in Planck units)
\begin{eqnarray}
\dot{M} \equiv \frac{dM}{dv} = -\frac{\alpha}{M^2},
\label{Mdot}
\end{eqnarray}
where $\alpha$ is a constant that has been numerically evaluated \cite{Page:1976df,Page:1976ki,Page:1977um,Page:1983ug,Page:2004xp} to be about $3.7474\times 10^{-5}$.

The apparent horizon is located at $r_{_{\mathrm{ApH}}} = 2 M(v)$, whereas the event horizon is generated by radially outgoing null geodesics, which obey
\begin{eqnarray}
\dot{r} \equiv \frac{dr}{dv} = \frac{1}{2} \left( 1 - \frac{2M(v)}{r} \right),
\label{null}
\end{eqnarray}
and are on the boundary of such null geodesics reaching out to future null infinity, instead of falling in to the singularity that is believed to be inside the black hole. For a smooth evaporation rate given by Eq.\ (\ref{Mdot}), the event horizon is given by the solution to Eq.\ (\ref{null}) such that it does not diverge exponentially far away from the apparent horizon in the future.  If we define $u \equiv 1 - r/(2M)$ and $p \equiv -4\dot{M}$ and assume that $n \equiv -d\ln{p}/d\ln{M}$ is constant, then one can show that the event horizon is at
\begin{eqnarray}
u&=&p + (n-2)p^2 + (n-1)(2n-5)p^3 \nonumber \\
&+& (6n^3-28n^2+37n-14)p^4 + O(p^5).
\label{adiabatic-horizon}
\end{eqnarray}
For a smooth Hawking evaporation into massless particles with $p \equiv -4\dot{M} = 4\alpha/M^2$, so that $n=2$, one finds that the event horizon is at
\begin{eqnarray}
r_{_\mathrm{EH}} = 2M[1-4\alpha/M^2+O(\alpha^3/M^6)].
\label{event-horizon}
\end{eqnarray}

For a general spherical metric, the covariant generalization of $d/dv$ along an outward null direction toward the future is $d/dv \equiv N^\alpha\partial/\partial x^\alpha$, with outward null vector $N^\alpha$ normalized so that $\dot{r} \equiv dr/dv \equiv N^\alpha r_{,\alpha} = (1/2)\nabla r\cdot\nabla r = (1/2)-M/r$.  Note that $dM/dv = -\alpha/M^2$ implies that $d^2(M^3)/dv^2 = 0$, but since $r_{_\mathrm{EH}} \approx 2M$, we have $d^2(r_{_\mathrm{EH}}^3)/dv^2 \approx 0$ as well.  Let us therefore define an {\it adiabatic horizon} at $r_{_\mathrm{AdH}}$ by the outer root of
\begin{eqnarray}
\frac{d^2}{dv^2}(r_{_\mathrm{AdH}}^3) \equiv N^\alpha\frac{\partial}{\partial x^\alpha}\left(N^\beta\frac{\partial}{\partial x^\beta}r_{_\mathrm{AdH}}^3\right) = 0.
\label{adia-horizon}
\end{eqnarray}
The location of the adiabatic horizon is very near where the event horizon would be if the future evolution of the latter followed the adiabatic mass evolution law of Eq.\ (\ref{Mdot}). One can show that $r_{_\mathrm{AdH}}$ is equivalent to the location where the gradient vector of $(1/4)\nabla(r^2)\cdot\nabla(r^2) = r^2\nabla r\cdot\nabla r \equiv r^2 - 2Mr$ (which, incidentally, defines $M$) is in the outward null direction, or $N^\alpha (r^2\nabla r\cdot\nabla r)_{,\alpha} = 0$, which gives $\dot{M} = (1/2) - (3/2)(M/r) + (M/r)^2$ and
\begin{eqnarray}
r_{_\mathrm{AdH}} \equiv \frac{4M}{3-\sqrt{1+16\dot{M}}}.
\label{adiabatic-horizon-formula}
\end{eqnarray}
When Eq.\ (\ref{Mdot}) holds, this expression agrees with Eq.\ (\ref{event-horizon}) to the order given.

\begin{figure}
\begin{center}
\includegraphics[scale=0.65]{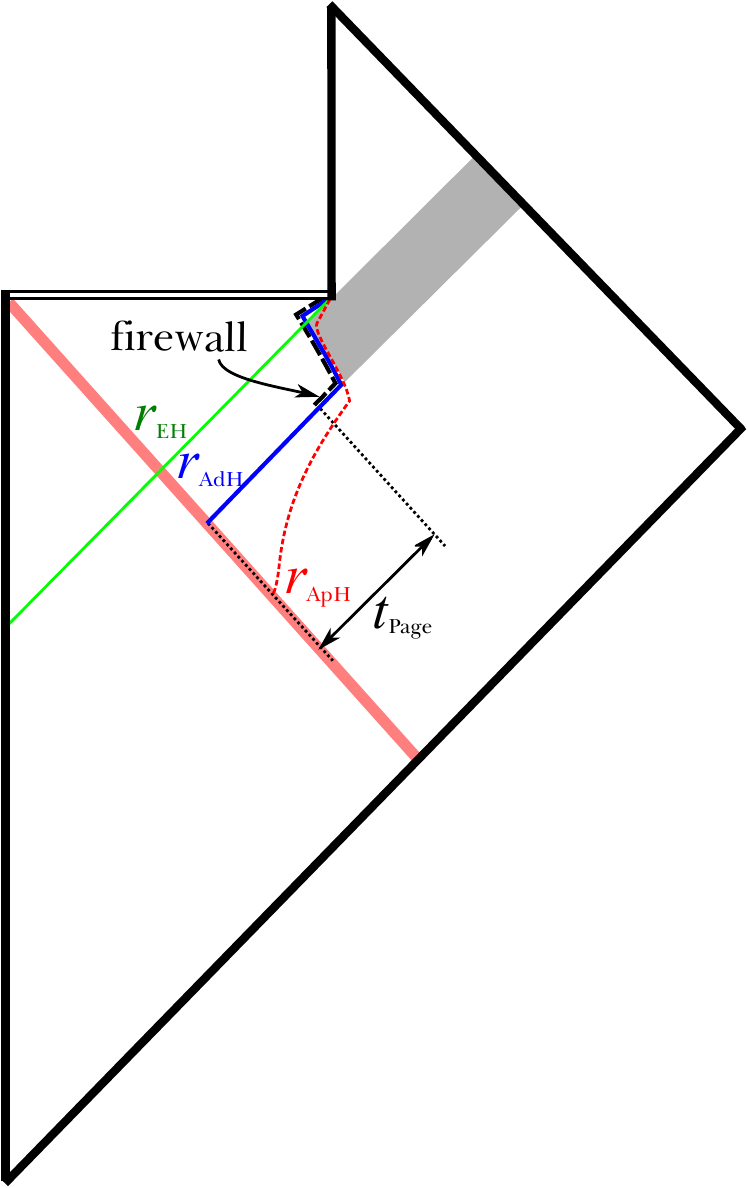}
\caption{\label{fig:pic}A conceptual Penrose diagram illustrating the formation of a Schwarzschild black hole from a collapsing null shell, and its subsequent Hawking evaporation. Here, the event horizon ($r_{_\text{EH}}$) has been shifted inward some distance from the adiabatic horizon ($r_{_\text{AdH}}$) due to a quantum fluctuation.  This renders the firewall (denoted by the dashed curve that appears after the Page time $t_{_\text{Page}}$) naked. The apparent horizon ($r_{_\text{ApH}}$) is also shown for comparison, but light rays can escape from inside it, since the black hole is shrinking.}
\end{center}
\end{figure}

We shall assume that the firewall, if it exists, is close to where the event horizon would be if the black hole evolved smoothly and adiabatically according to Eq.\ (\ref{Mdot}).  However, the actual event horizon depends on the future evolution of the spacetime, and not just on that of its past.  Therefore, quantum fluctuations in the future spacetime can lead the event horizon to deviate significantly from the unperturbed adiabatic horizon. If the mass loss rate exceeds the adiabatic formula, then the event horizon will be inside the adiabatic horizon. As a result a firewall located at the adiabatic horizon would become naked, visible from future null infinity. See Fig.\ (\ref{fig:pic}) for a diagrammatic explanation.

From Eq.\ (\ref{null}), one can write the mass $M = M(v)$ in the Vaidya metric in terms of the event horizon radius $r = r(v) \equiv r_{_{\mathrm{EH}}}(v)$ as
\begin{eqnarray}
M = \frac{1}{2} r - r \dot{r}.
\label{mass-radius}
\end{eqnarray}
Let $M_1, r_1$ and $M_2, r_2$ be the unperturbed mass and radius of the black hole and their fluctuations, respectively, with total mass $M=M_1+M_2$ and event horizon radius $r=r_1+r_2$.\footnote{Note that we are comparing the true event horizon at $r = r_1 + r_2$ with where it would have been, at $r_1$, if there were no perturbation $r_2$, but this is not the same as what Eq.\ (\ref{adiabatic-horizon-formula}) would define as the ``adiabatic horizon'' when the perturbation is included.  The adiabatic horizon would be very near the event horizon when the mass loss rate has a smooth form such as that given by Eq.\ (\ref{Mdot}), but for a significant perturbation $\dot{M}_2$ in the mass loss rate, the adiabatic horizon need not be near either the event horizon at $r$ or the unperturbed horizon at $r_1$.  We hence emphasize that the unperturbed horizon should not be confused with the adiabatic horizon once the mass perturbation $M_2$ becomes significant.}  Now suppose that the unperturbed mass loss would give $M = M_1 = M_1(v) = (1/2)r_1 - r_1 \dot{r}_1$, such that $\dot{M}_1 \approx -\alpha/M_1^2$, and that quantum fluctuations $M_2 = M_2(v)$ and $r_2 = r_2(v)$ are small compared with the total mass and the event horizon radius, respectively.  Then
\begin{eqnarray}\label{approx}
M &=& M_1 + M_2 = \frac{1}{2} r - r \dot{r} \nonumber \\
&=& \frac{1}{2} (r_1 + r_2) - (r_1 + r_2)(\dot{r}_1 + \dot{r}_2) \nonumber \\
&\approx& M_1 + \frac{1}{2} r_2 - r_1 \dot{r}_2.
\label{mass-radius-fluctuations}
\end{eqnarray}
For simplicity we are making the highly idealized assumption that even with quantum fluctuations, the metric remains spherically symmetric and Vaidya near the event horizon, though this is not crucial for our argument.

Now for some particular advanced time $v = v_0$, let us ignore quantum fluctuations before this time, so that $M_2(v) = 0$ for $v < v_0$, and let us define the constant $M_0 = M(v_0) = M_1(v_0)$.  To leading order in $M_0 \gg 1$ and $|v-v_0| \ll M_0^3$, the fractional decay of the black hole over the advanced time $v - v_0$ is small, and the negative of the coefficient of $\dot{r}_2$ in Eq.\ (\ref{mass-radius-fluctuations}) may be written as $r_1 \approx 2M_1 \approx 2M_0$.  Then Eq.\ (\ref{mass-radius-fluctuations}) gives $(1/2) r_2 - 2M_0\dot{r}_2 \approx M_2(v)$. The solution of this differential equation that is void of an exponentially growing departure of the event horizon $r(v) = r_1 + r_2$ from the unperturbed horizon $r_1(v)$ at late times is
\begin{eqnarray}
r_2 \approx \exp{\left(\frac{v-v_0}{4M_0}\right)}
\int_v^\infty dv' \frac{M_2(v')}{2M_0} \exp{\left(\frac{v_0-v'}{4M_0}\right)}.
\label{radius-fluctuations}
\end{eqnarray}

Since the unperturbed evolution gives $\dot{M}_1 \approx -\alpha/M_0^2$ for $M_0 \gg 1$ and $|v-v_0| \ll M_0^3$, let us consider a quantum mass fluctuation that gives, with $\theta(v-v_0)$ the Heaviside step function,
\begin{eqnarray}
\dot{M}_2 = -\theta(v-v_0)\frac{\alpha\beta}{M_0^2} 
\exp{\left(-\frac{\gamma(v-v_0)}{4M_0}\right)},
\label{energy-emission-rate-fluctuations}
\end{eqnarray}
which has two new constant parameters, namely $\beta$ for how large the quantum fluctuation in the energy emission rate is relative to the unperturbed emission rate $-\alpha/M^2$ (with $\beta$ assumed to be positive so that the quantum fluctuation increases the emission rate above the unperturbed value), and $\gamma$ for how fast the quantum fluctuation in the energy emission rate decays over an advanced time of $4M_0$ (the inverse of the surface gravity $\kappa$ of the black hole).  Then with $M_2(v) = 0$ for $v < v_0$, one gets
\begin{eqnarray}
M_2 \approx -\theta(v-v_0)\frac{4\alpha\beta}{\gamma M_0} 
\left[1 - \exp{\left(-\frac{\gamma(v-v_0)}{4M_0}\right)}\right].
\label{mass-fluctuations}
\end{eqnarray}
Plugging this back into Eq.\ (\ref{radius-fluctuations}) then gives
\begin{eqnarray}
\!\!\!\!\!\!&r_2&\approx - \theta(v_0-v)\frac{8\alpha \beta}{(1+\gamma)M_0} \exp{\left(\frac{v-v_0}{4M_0}\right)} \nonumber \\
-\!\!\!\!&\theta&\!\!\!\!\!(v\!-\!v_0)\frac{8\alpha \beta}{\gamma(1\!+\!\gamma)M_0} 
\!\left[\!1\!+\!\gamma\!-\!
\exp{\!\left(-\frac{\gamma(v\!-\!v_0)}{4M_0}\!\right)}\!\right]\!\!.
\label{particular-radius-fluctuations}
\end{eqnarray}

This particular form of the emission rate fluctuation implies that the total mass fluctuation from the unperturbed evolution is $M_2(\infty) = - 4\alpha \beta/(\gamma M_0)$.  Then the radial fluctuation in the event horizon radius at the advanced time $v = v_0$, when $-r_2(v)$ has its maximum value, is
\begin{eqnarray}
r_2(v_0) \approx \frac{2\gamma}{1+\gamma} M_2(\infty).
\label{maximum-particular-radius-fluctuations}
\end{eqnarray}
This means that if the quantum fluctuation in the energy emission rate is very short compared with $4 M_0$ (decaying rapidly in comparison with the surface gravity of the black hole), so that $\gamma \gg 1$, then $r_2(v_0) \approx 2 M_2(\infty)$, twice the total mass fluctuation.  However, we shall just assume that $\gamma$ is of the order of unity and hence get $r_2(v_0) \sim M_2(\infty)$ as an order-of-magnitude relation.  Note that the reduction in the radius of the event horizon at $v = v_0$, where the fluctuation in the mass emission rate starts, occurs {\it before} there is any decrease in the mass below the unperturbed value $M_1(v)$, because the location of the ``teleological'' event horizon is defined by the future evolution of the spacetime.

Note that $r_1 \sim M_1 \sim M_0$, $r_2 \sim 1/M_0$, $\dot{r}_1 \sim 1/M_0$, and $\dot{r}_2 \sim 1/M_0^2$. This is consistent with the approximations made in Eq.\ (\ref{approx}) to drop the terms $r_2\dot{r}_2$ and $r_2\dot{r}_1$. 

Therefore, if the putative firewall occurs at a location determined purely causally by the past behavior of the spacetime, and is sufficiently near where the event horizon would be under unperturbed adiabatic emission thereafter (say near the adiabatic horizon), then quantum fluctuations, at later advanced times that reduce the mass of the hole below that given by the unperturbed evolution, would move the actual event horizon inward (even before quantum fluctuations in the mass emission rate begin), so that the event horizon becomes inside the location of the putative firewall.  That is, quantum fluctuations that increase the mass emission rate render such a firewall naked, visible to the external universe.

One possible objection to this conclusion is that for $\alpha$, $\beta$, and $\gamma$ all of the order of unity, the inward shift in the event horizon is by a change of radius, $r_2$, of the order of $1/M$, so that the proper distance from the putative firewall near $r = r_1$ to the event horizon at $r = r_1 + r_2$, in the frame of the timelike firewall surface outside the event horizon, is of the order of the Planck length.  The proper acceleration of an observer that stays of the order of the Planck length outside the event horizon would be of the order of the Planck acceleration, giving an Unruh temperature of the order of the Planck temperature.  One might object that quantum gravity effects at such extreme accelerations would make a naked firewall in practice indistinguishable from a firewall at or inside the event horizon.

However, for a black hole of huge initial entropy $S \gg 1$ that emits roughly $S$ particles during its Hawking evaporation, there are a large number of roughly $S$ approximately independent chances for the proper distance fluctuation of the event horizon inside the firewall to reach a large value, say $L \gg 1$, so that the probability at any one time needs only be $P(L) \sim 1/S$.  For a large fluctuation $L$, the most probable way to produce it at $v = v_0$, when the Hawking temperature is $T_0 = 1/(8\pi M_0)$, is to have thereafter the radiation be locally thermal with a time-dependent temperature $T(v) = T_0(z+1)/[z+1-z e^{-(v-v_0)/(4M_0)}]$ with a constant parameter $z = [T(v_0)-T_0]/T_0 \gg 1$ chosen to give the desired $L = [8M_0(-r_2(v_0))]^{1/2}$.  The probability of this fluctuation then works out to be $P(L) \sim \exp{[-(\pi/2)L^2]}$.  Setting this to be $\sim 1/S$ then gives the most probable largest value of the fluctuation as $L \approx [(2/\pi)\ln{S}]^{1/2}$, which is arbitrarily large for arbitrarily large $S$.  Therefore, arbitrarily large black holes can have the event horizon fluctuate an arbitrarily great distance inside a firewall whose location is determined causally.  Hence, the firewall of an arbitrarily large black hole will with high probability become highly naked, observable without encountering quantum gravity effects (other than what quantum gravity effects are supposed to lead to the existence of the firewall itself).

Therefore the firewall is \emph{not} hidden in the region with Planckian local temperature; its presence would truly be at odds with expectations from general relativity and ordinary quantum field theory.
More specifically, being in the exterior of the event horizon means that the firewall could potentially influence the exterior spacetime, so that even observers who do not fall into the black hole could have a fiery experience.  In addition, the presence of a firewall well outside the event horizon could affect the spectrum of the Hawking radiation, which means that the presence of a firewall could be inferred even by asymptotic observers. Such a ``naked firewall,'' i.e., a firewall far outside the event horizon, is therefore problematic, and giving up the no-drama assumption no longer seems like a palatable ``most conservative'' option.  



A natural interpretation is that if there is a firewall, then it should affect not only the interior geometry of the black hole, but also the asymptotic future. The former would ``only'' violate general relativity for a free-falling observer, while the latter would violate the semi-classical quantum field theory for an asymptotic observer \cite{dh1,dh2}. Thus the firewall solution can hardly be considered as conservative.

We gratefully acknowledge the generous support and the stimulating environment provided by the Yukawa Institute for Theoretical Physics through its Molecule-type Workshop on ``Black Hole Information Loss Paradox" (YITP-T-15-01), where the concept of this paper was formulated. In addition, the work of PC is supported in part by Taiwan's National Center for Theoretical Sciences (NCTS) and Taiwan's
Ministry of Science and Technology (MOST), of DNP by the Natural Sciences and Engineering Council of Canada, of MS by MEXT KAKENHI Grant Number 15H05888, and of DY by the Leung Center for Cosmology and Particle Astrophysics (LeCosPA) of National Taiwan University (103R4000). YCO thanks Nordita for various travel supports.

\end{document}